\documentclass[twocolumn,showpacs]{revtex4}
\usepackage{amssymb}
\usepackage{amsmath}
\usepackage{epsfig}
\begin{document}

\title{Quantum phase transition in a three-level atom-molecule system}
\author{Sheng-Chang Li$^{1}$}
\email{scli@mail.xjtu.edu.cn}
\author{Li-Bin Fu$^{2,3}$}
\author{Fu-Li Li$^{1}$}

\affiliation{1.School of Science, Xi'an Jiaotong University, Xi'an
710049, China\\2.National Key Laboratory of Science and Technology
on Computation Physics, Institute of Applied Physics and
Computational Mathematics, Beijing 100088, China
\\3.Center for Applied Physics and Technology, Peking University, Beijing 100084,
China}

\begin{abstract}
We adopt a three-level bosonic model to investigate the quantum
phase transition in an ultracold atom-molecule conversion system
which includes one atomic mode and two molecular modes. Through
thoroughly exploring the properties of energy level structure,
fidelity, and adiabatical geometric phase, we confirm that the
system exists a second-order phase transition from an atom-molecule
mixture phase to a pure molecule phase. We give the explicit
expression of the critical point and obtain two scaling laws to
characterize this transition. In particular we find that both the
critical exponents and the behaviors of ground-state geometric phase
change obviously in contrast to a similar two-level model. Our
analytical calculations show that the ground-state geometric phase
jumps from zero to $\pi/3$ at the critical point. This discontinuous
behavior has been checked by numerical simulations and it can be
used to identify the phase transition in the system.
\end{abstract}

\pacs{05.30.Rt, 67.85.Hj, 03.65.Vf}

\maketitle

\section{Introduction}

Quantum phase transition (QPT) is one of the most important concepts
in the many-body quantum theory. As a central fundamental transition
phenomenon at the temperature of absolute zero, it describes an
abrupt change in the ground state of a many-body system due to its
quantum fluctuations \cite{sachdevs2003,sondhisl1997}. The
experimental observation of a QPT from a superfluid (SF) to a Mott
insulator (MI) in a gas of ultracold atoms \cite{greinerm2002}
inspired great interest in investigating clean, highly controllable,
and strong correlated bosonic systems \cite{tilahund2011}. Actually,
the ultracold atomic gases \cite{pitaevskiil2003} have become an
ideal platform to study many-body physics because of their enormous
applications and the advanced experimental techniques available in
the fields of atomic and optical physics \cite{ruseckasj2005}.

In recent years, a remarkable development in the aforementioned
filed is to convert ultracold atoms to molecules via Feshbach
resonance \cite{Donleyea2002,xuk2003,Herbigj2003} or
photoassociation \cite{Wynarr2000,Romt2004,whinklerk2005}
techniques. Compared with the fermionic model in this kind of
systems, the bosonic model is of interest for theoretically
exploring the QPTs. When both ultracold atoms and molecules are
bosons, the systems possess a few degrees of freedom and a large
particle number which can greatly simplify the calculations. On the
other hand, the atom-molecule conversion systems can be well
described by a mean-field theory when the particle number is large
enough and this treatment will lead to nonlinearity. These features
have stimulated much efforts to study the adiabatic evolution
\cite{Gaubatzu1988,Kuklinskijr1989,Pazye2005,linghy2007,Shapiroea2007,mengsy08},
geometric phase \cite{fulb2010,wub2011}, and phase transition
\cite{santosg10,lisc11a} of the systems. Instead of the traditional
approaches for describing a QTP (i.e., using the concepts of order
parameter and symmetry breaking), very recently Santos {\it el al}.
adopted the concepts of entanglement and fidelity to investigate the
QPT in a two-level bosonic atom-molecule system \cite{santosg10}.
Motivated by this work we discussed same problems from the
perspectives of scaling laws and Berry curvature \cite{lisc11a}.
However, the connection between the mean-field Berry phase and the
phase transition in this type of bosonic model, and the properties
of the QPT in a three-level atom-molecule system are still
unresolved, which call for further theoretical considerations.

As a continuous work, in this paper we investigate the quantum phase
transition in an ultracold atom-molecule conversion system by
adopting a $\Lambda$-type three-level bosonic model. Based on this
model, we first discuss the structure of quantum energy levels and
analyze the properties of ground state. In order to compare the
phase transition properties with a similar two-level model
\cite{santosg10}, we study the energy gap, the ground-state
fidelity, and mean-field geometric phase. We illustrate that, when
the ratio of the coupling strength between two molecular modes to
that between atomic mode and the upper molecular mode exceeds a
critical value, a similar QPT from a mixed atom-molecule phase to a
pure molecular phase is also observed in our system. To characterize
this transition we obtain the analytical expression of the critical
point by using the mean-field approach and derive two critical
exponents via numerically studying scaling laws. In particular we
calculate the ground-state geometric phase and find its
discontinuous behavior at the phase transition point.

Our paper is organized as follow: In Sec. II we give the
second-quantized model and its mean-field description. In Sec. III
we explore the properties of energy levels and ground states. In
Sec. IV we choose the characteristic scaling law, fidelity, and
adiabatic geometric phase to describe the QPT. Section V presents
our conclusion.

\section{Three-level model and mean-field description}

\begin{figure}[th]
\begin{center}
\rotatebox{0}{\resizebox *{5.5cm}{4.cm} {\includegraphics
{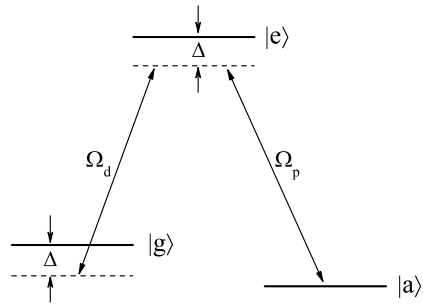}}}
\end{center}
\caption{Schematic diagram of a three-level atom-molecule conversion
model coupled by two laser pulses $\Omega_d$ and $\Omega_p$. The
stable atomic state, the molecular ground state, and the molecular
excited state are denoted by $|\mathrm{a}\rangle$,
$|\mathrm{g}\rangle$, and $|\mathrm{e}\rangle$, respectively.
$\Delta$ measures the detuning of the pump laser $\Omega_p$ with
respect to the transition from $|\mathrm{a}\rangle$ to
$|\mathrm{e}\rangle$.}\label{fig1}
\end{figure}

The system we consider here is illustrated schematically in Fig.
\ref{fig1}. It describes the process of creating ultracold diatomic
molecules from bosonic condensed atoms, which constitutes a
$\Lambda$-type three-level model. In the three-mode description,
each mode $|\alpha\rangle$ ($\alpha=a,g$, and $e$ respectively
represent the atomic mode, the ground-state molecular mode, and the
excited-state molecular mode) is associated with an annihilation
operator $\hat{\beta}$ ($\beta=a,b_g$, and $b_e$) due to the basic
assumption that the spatial wavefunctions for these modes are fixed.
By setting the energy of atomic mode as zero, the Hamiltonian of the
system takes the following second-quantized form with $\hbar=1$
\cite{wub2011}:
\begin{align}\label{H1}
\hat{H}_S=&\omega_e\hat{b}_e^\dagger\hat{b}_e+\omega_g\hat{b}_g^\dagger\hat{b}_g
+\notag\\&\Omega_de^{i\nu_dt}\hat{b}_e^\dagger\hat{b}_g+\frac{\Omega_pe^{-i\nu_pt}}{\sqrt{N}}\hat{b}_e^\dagger
\hat{a}\hat{a}+\mathrm{H.c.},
\end{align}
where the abbreviation $\mathrm{H.c.}$ denotes the operation of
Hermitian conjugate. $\nu_d$ and $\nu_p$ are the frequencies of two
laser pulses $\Omega_d$ and $\Omega_p$, respectively. The
frequencies $\omega_g$ and $\omega_e$ measure the molecular ground
state and excited state energies, respectively. The total atom
number $N=N_a+2(N_g+N_e)$ with $N_a=\hat{a}^\dagger\hat{a}$,
$N_g=\hat{b}_g^\dagger\hat{b}_g$, and
$N_e=\hat{b}_e^\dagger\hat{b}_e$, commutes with the Hamiltonian
(\ref{H1}) and is therefore conserved. Notice that the laser pulse
parameter $\Omega_p$ can be complex. To achieve this one can split
the laser pulse into two beams and then recombine and focus them on
the system. As a result, we can express the complex parameter
$\Omega_p$ as $\Omega_p=\xi_1+\xi_2e^{-i\varphi}$ with $\xi_1$ and
$\xi_2$ being real numbers, where the phase factor $\varphi$ is
determined by the difference of optical paths  between the two laser
beams \cite{wub2011}.

For convenience, we rewrite the above Schr\"{o}dinger picture
Hamiltonian as, $\hat{H}_S=\hat{H}_0+\hat{H}_1$, where
\begin{align}
\hat{H}_0=&\nu_p\hat{b}_e^\dagger\hat{b}_e+(\nu_p-\nu_d)\hat{b}_g^\dagger\hat{b}_g,\\
\hat{H}_1=&(\omega_e-\nu_p)\hat{b}_e^\dagger\hat{b}_e+(\omega_g-\nu_p+\nu_d)
\hat{b}_g^\dagger\hat{b}_g+\notag\\&\Omega_de^{i\nu_dt}\hat{b}_e^\dagger\hat{b}_g+{\Omega_pe^{-i\nu_pt}\over\sqrt{N}}\hat{b}_e^\dagger
\hat{a}\hat{a}+\mathrm{H.c.},
\end{align}
then we choose $\omega_e=\omega_g+\nu_d$ and apply the interaction
picture, i.e., $\hat{H}_I=e^{i\hat{H}_0t}\hat{H}_1e^{-i\hat{H}_0t}$,
the Hamiltonian finally becomes
\begin{align}\label{H2}
\hat{H}_I=\Delta(\hat{b}_e^\dagger\hat{b}_e+\hat{b}_g^\dagger\hat{b}_g)+z\hat{b}_e^\dagger\hat{b}_g+{\rho
e^{-i\phi}\over\sqrt{N}}\hat{b}_e^\dagger
\hat{a}\hat{a}+\mathrm{H.c.},
\end{align}
where the new parameters $\Delta=\omega_e-\nu_p$, $z=\Omega_d$,
$\rho=|\Omega_p|$, and $\phi=\arg(\Omega_p)$ have been introduced.

To complement the quantum description and gain insights into the
existence of a QPT in our model, we adopt a semiclassical
description of the system by following the usual mean-field
approach, which has been proven to be a powerful tool for studying
ultracold atoms and Bose-Einstein condensates (BECs). In the
semiclassical limit $N\rightarrow\infty$, the quantum model becomes
classical and one can replace the operator $\hat{\beta}$ with a
corresponding complex number $\beta$ ($\beta=a,b_g,b_e$), i.e.,
$\mathcal{H}=\Delta(|{b}_e|^2+|{b}_g|^2)+z({b}_e^\ast{b}_g+{b}_g^\ast{b}_e)+\rho
[e^{-i\phi}{b}_e^\ast{a}^2+e^{i\phi}{b}_e(a^\ast)^2]$. By using the
equations $id{\beta}/dt={\partial\mathcal{H}}/{\partial\beta^\ast}$,
we can obtain the following Schr\"{o}dinger equations together with
the normalization condition $|a|^2+2(|b_g|^2+|b_e|^2)=1$ to govern
the dynamical behaviors of the system,
\begin{align}\label{Hmf}
i\frac{d}{dt}|\psi\rangle=H_{mf}|\psi\rangle,
\end{align}
where
\begin{align}
H_{mf}=\left(
           \begin{array}{ccc}
             0 & 0 & 2\rho{e}^{i\phi}{a}^\ast \\
             0 & \Delta & z \\
             \rho{e}^{-i\phi}{a} & z & \Delta \\
           \end{array}
         \right),
\end{align}
and $|\psi\rangle=(a,b_g,b_e)^T$. It is worth emphasizing that,
although the collisions between ultracold particles have been
neglected in our model, the nonlinearity also arises from the
mean-field treatment for the fact of two atoms to form one molecule.
Mathematically, the mean-field Hamiltonian $H_{mf}$ is a function of
the instantaneous wave function as well as its conjugate. It is a
nonhermitian matrix and it is invariant under the following
transformation \cite{mengsy08}:
\begin{align}
|\psi\rangle\rightarrow
U_s|\psi\rangle={e}^{i\Theta(\theta)}|\psi\rangle={e}^{i{\left(
\begin{smallmatrix}
\theta & 0 & 0 \\
0 & 2\theta & 0 \\
0 & 0 & 2\theta \\
\end{smallmatrix}
\right)}}|\psi\rangle.
\end{align}
The lack of $U(1)$ gauge transformation is a particular interesting
point of the above mean-field model, which may lead to some new
properties of the system. In the subsequent sections, based on
models (\ref{H2}) and (\ref{Hmf}), we will discuss the QPT in the
system both from the fully quantum perspective and the mean-field
perspective.

\section{Energy levels and Ground states}

Taking advantage of the fact of $N$ is conserved, one can
diagonalize the quantum Hamiltonian $\hat{H}_I$. For simplicity,
hereafter we assume that $N$ takes an even constant value, then the
Hilbert space of the $N$-particle system can reduce to
$\tfrac{1}{2}(\tfrac{N}{2}+1)(\tfrac{N}{2}+2)$ dimension in the Fock
basis, i.e.,
$|n_a\rangle|n_g\rangle|n_e\rangle=(n_a!n_g!n_e!)^{-1/2}(\hat{a}^\dagger)^{n_a}
(\hat{b}_g^\dagger)^{n_g}(\hat{b}_e^\dagger)^{n_e}|0\rangle$ with
$|0\rangle$ being the vacuum state, where
$n_g=0,1,\cdots,\tfrac{N}{2}$, $n_e=0,1,\cdots,\tfrac{N}{2}-n_g$,
and $n_a=N-2(n_g+n_e)$ represent the populations of particle in
states $|b_g\rangle$, $|b_e\rangle$, and $|a\rangle$, respectively.
\begin{figure}[t]
\begin{center}
\rotatebox{0}{\resizebox *{8.5cm}{7.5cm} {\includegraphics
{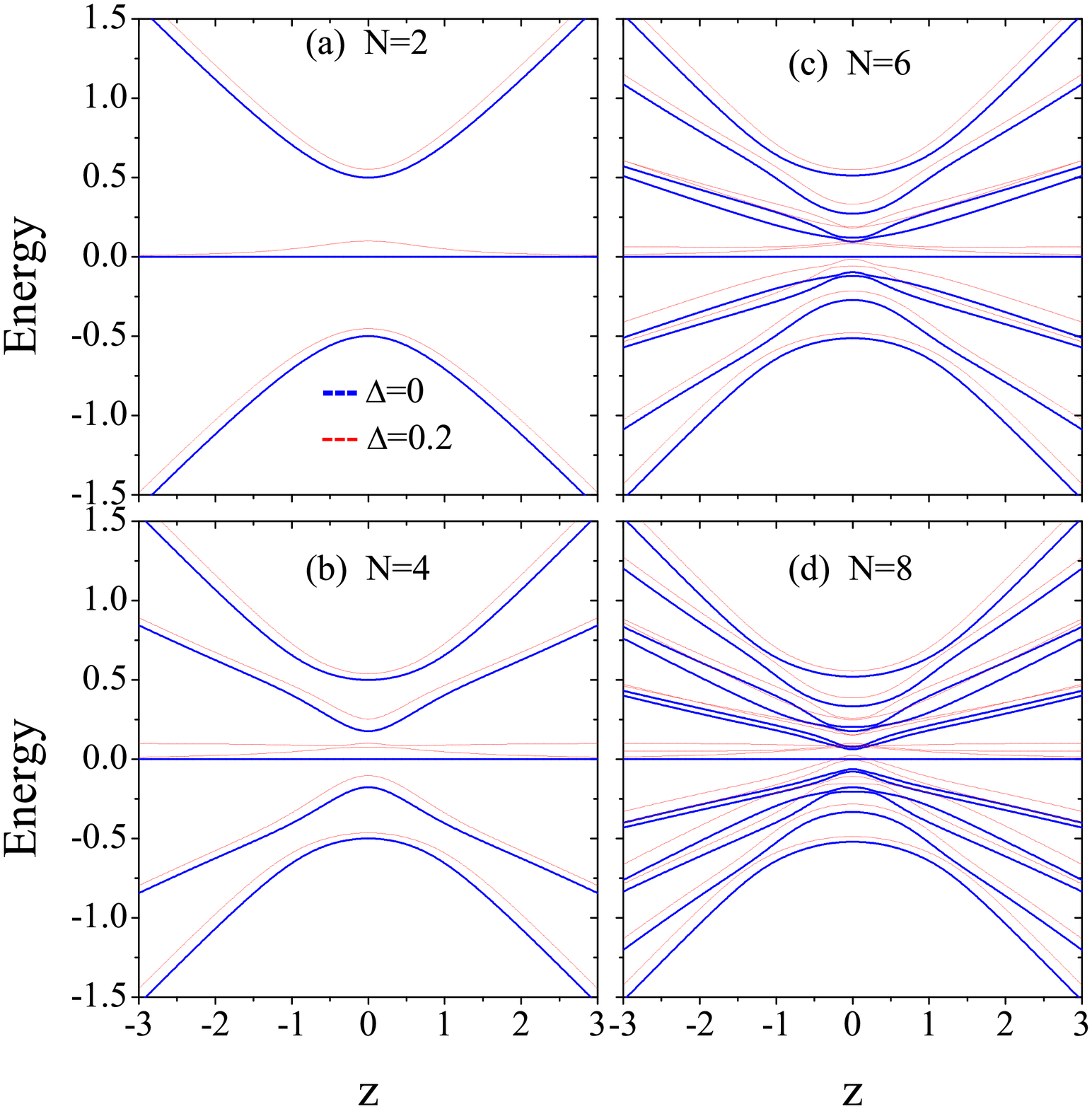}}}
\end{center}
\caption{(Color online) Quantum energy levels for different particle
numbers: (a) $N=2$, (b) $N=4$, (c) $N=6$, and (d) $N=8$. The thick
and thin lines denote the cases $\Delta=0$ and $\Delta=0.2$,
respectively. The energies are shown in units of $N\rho$ while the
parameters $z$ and $\Delta$ are rescaled by $\rho$.}\label{fig2}
\end{figure}
By directly diagonalizing the Hamiltonian matrix with a fixed $N$,
we obtain the eigen-energy levels and the ground states of the
system as shown in Figs. \ref{fig2} and \ref{fig3}, respectively.

It is well known that a typical $\Lambda$-type three-level system
supports dark-state solutions with zero eigenvalue
\cite{puh07,linghy04,linghy07}. This type of state can result in a
phenomenon known as coherent population trapping (CPT). For our
system, when $\Delta=0$, from Figs. \ref{fig2}(b)-\ref{fig2}(d) we
see that the energy levels with energy value being zero are
degenerate while other nonzero-energy levels are nondegenerate and
are symmetrically distributed in both sides of the center level.
This symmetrical energy structure is determined by the symmetry of
the Hamiltonian $\hat{H}_I$ with $\phi=0$, i.e., the change of
variables $(z,\rho)\rightarrow -(z,\rho)$ is equivalent to the
unitary transformation $\hat{b}_e\rightarrow-\hat{b}_e$. In this
case, the degeneracy of the zero-energy level (i.e., $d$) is given
by
\begin{align}
d=\lceil(\frac{N}{2}+1)/2\rceil=\frac{1}{4}(N-\mathrm{Mod}[N,4])+1,
\end{align}
the symbol $\lceil~\rceil$ stands for the ceiling function which
maps a real number to the smallest following integer.

Notice that, if the parameter $\Delta$ has a perturbation, the above
mentioned symmetry of the system with $\Delta=0$ will be broken.
This leads to the energy levels shift and the zero-energy level
splitting. For example, when $\Delta=0.2$ and $N=8$ [see Fig.
\ref{fig1}(d)], all energy levels have been pushed up and the
zero-energy level has split into three nonzero-energy levels. For
different $N$, the maximum number of energy levels should be
$(\tfrac{N}{2}+1)(\tfrac{N}{2}+2)/2$.
\begin{figure}[t]
\begin{center}
\rotatebox{0}{\resizebox *{8.5cm}{6cm} {\includegraphics
{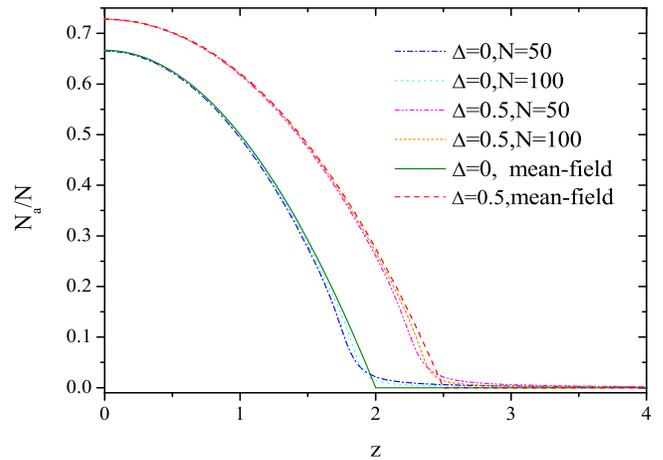}}}
\end{center}
\caption{(Color online) Atomic fraction in the ground state versus
the parameter $z$ with $\Delta=0$ and $\Delta=0.5$. Both $z$ and
$\Delta$ are rescaled by $\rho$.}\label{fig3}
\end{figure}

Now we discuss the ground-state properties which are closely
associated with the QPT in the system. On one hand, by diagonalizing
the Hamiltonian $\hat{H}_I$ numerically, both for $\Delta=0$ and for
$\Delta\neq 0$, we have calculated the ground states with different
total atom numbers. The results for atomic population fraction
(i.e., $N_a/N$) in the ground state are demonstrated in Fig.
\ref{fig3}. We find that the atomic fraction in the ground state
decreases and gradually approaches to zero as the ratio of the
coupling strength between two molecular modes to that between atom
mode and the molecular mode increases. On the other hand, we study
the ground state from the mean-field perspective. Based on the model
(\ref{Hmf}), we solve the mean-field ground state from the
eigen-equation
$H_{mf}(\bar{a}^\ast,\bar{a})|\bar{\psi}\rangle=\Theta(\mu)|\bar{\psi}\rangle$
with $\mu$ being the chemical potential for atoms and
$|\bar{\psi}\rangle$ being the eigenstate. For $\Delta=0$, $z\geq
0$, and $\rho>0$, we obtain the eigenvalue and the corresponding
eigenfunction for the ground state as follows:
\begin{align}
\mu_0=&\left\{
      \begin{array}{cc}
        -{z\over 2}, & z>2\rho, \\
        -{(z^2-2\rho^2)\sqrt{z^2+8\rho^2}\over 4\sqrt{3}\rho^2}, & z<2\rho; \\
      \end{array}
    \right.\\
|\bar{\psi_0}\rangle=& \left\{
  \begin{array}{cc}
    \left(
       \begin{array}{c}
         0 \\
         {1\over 2} \\
         -{1\over 2} \\
       \end{array}
     \right),
     & z>2\rho, \\
    \left(
       \begin{array}{c}
         {\sqrt{4-z^2/\rho^2}\over\sqrt{6}} \\
         {z\over 4\rho}e^{-i\phi} \\
         -{\sqrt{z^2+8\rho^2}\over 4\sqrt{3}\rho}e^{-i\phi}  \\
       \end{array}
     \right),
     & z<2\rho. \\
  \end{array}
\right.\label{mfgs}
\end{align}
\begin{figure}[t]
\begin{center}
\rotatebox{0}{\resizebox *{8.5cm}{5.cm} {\includegraphics
{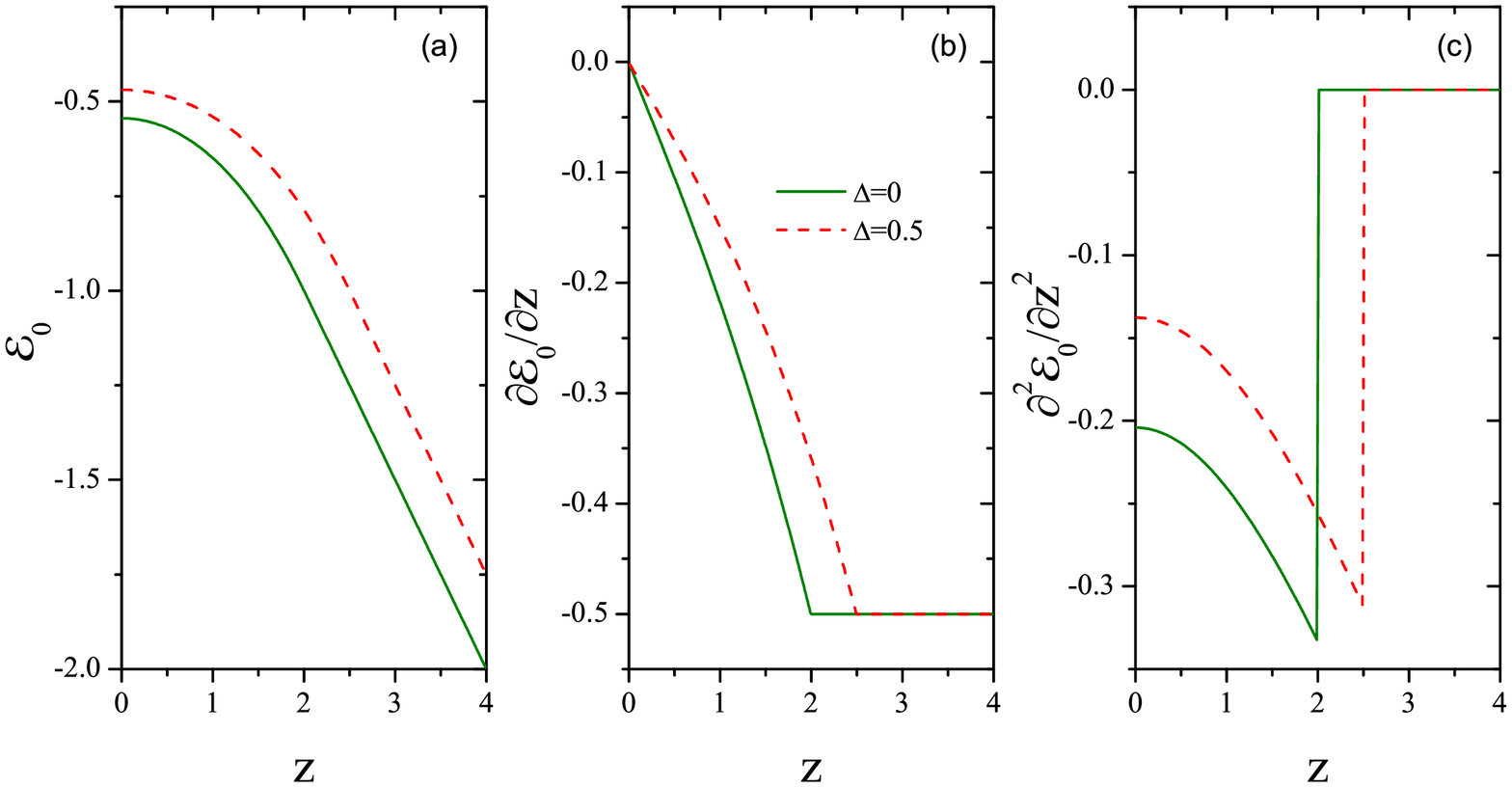}}}
\end{center}
\caption{(Color online) The energy of the mean-field ground state
(a) and its first (b) and second (c) derivatives with respect to
coupling $z$. The parameters $z$ and $\Delta$ are rescaled by
$\rho$.}\label{fig3a}
\end{figure}
For $\Delta\neq 0$, although the solutions to ground state can also
be obtained analytically, the expressions are generally too messy to
be instructive. We therefore simply display the results in Fig.
\ref{fig3}. From Fig. \ref{fig3} we find that, both for $\Delta=0$
and for $\Delta\neq 0$, the results for the quantum model (\ref{H2})
will tend to the analytical mean-field results with increasing the
total particle number $N$.

It must be mentioned that, for our nonlinear system (\ref{Hmf}), the
classical energy $\mathcal{E}$ does not equal the chemical
potential, and the relation between them is
$\mathcal{E}=\mu\pm\rho|\bar{b_e}||\bar{a}|^2$. We have calculated
the ground-state energy $\mathcal{E}_0$ analytically and find its
second derivative possesses a discontinuity at a critical point
$z_c=2\rho+\Delta$ as demonstrated in Fig. \ref{fig3a}. This
divergence behavior implies that the system exists a second-order
phase transition in the thermodynamic limit. When $z<z_c$ the system
is in an atom-molecule mixture phase (i.e., $|\bar{a}|^2>0$) and
when $z>z_c$ the system is in a pure molecule phase where
$|\bar{a}|^2=0$. In the mixture phase, the asymptotic behavior of
the ground state in the vicinity of the critical point with
$\Delta=0$ is given by the variation of the parameter
$s_0=|\bar{a}|^2$, i.e.,
\begin{align}
s_0|_{z\rightarrow z_c}=\frac{1}{6}[4 - z_c(2 z- z_c)].
\end{align}

\section{Quantum phase transition}

The previous calculations and analysis have demonstrated that the
process of converting ultracold atoms to homonuclear diatomic
molecules in bosonic system is a QPT which differs from the
well-known BCS-BEC crossover phenomena in fermionic systems
\cite{regalca2004,linksj2003}. Subsequently, we will describe and
characterize this phase transition from different perspectives.

\subsection{Scaling laws}

\begin{figure}[t]
\begin{center}
\rotatebox{0}{\resizebox*{8.5cm}{6cm} {\includegraphics {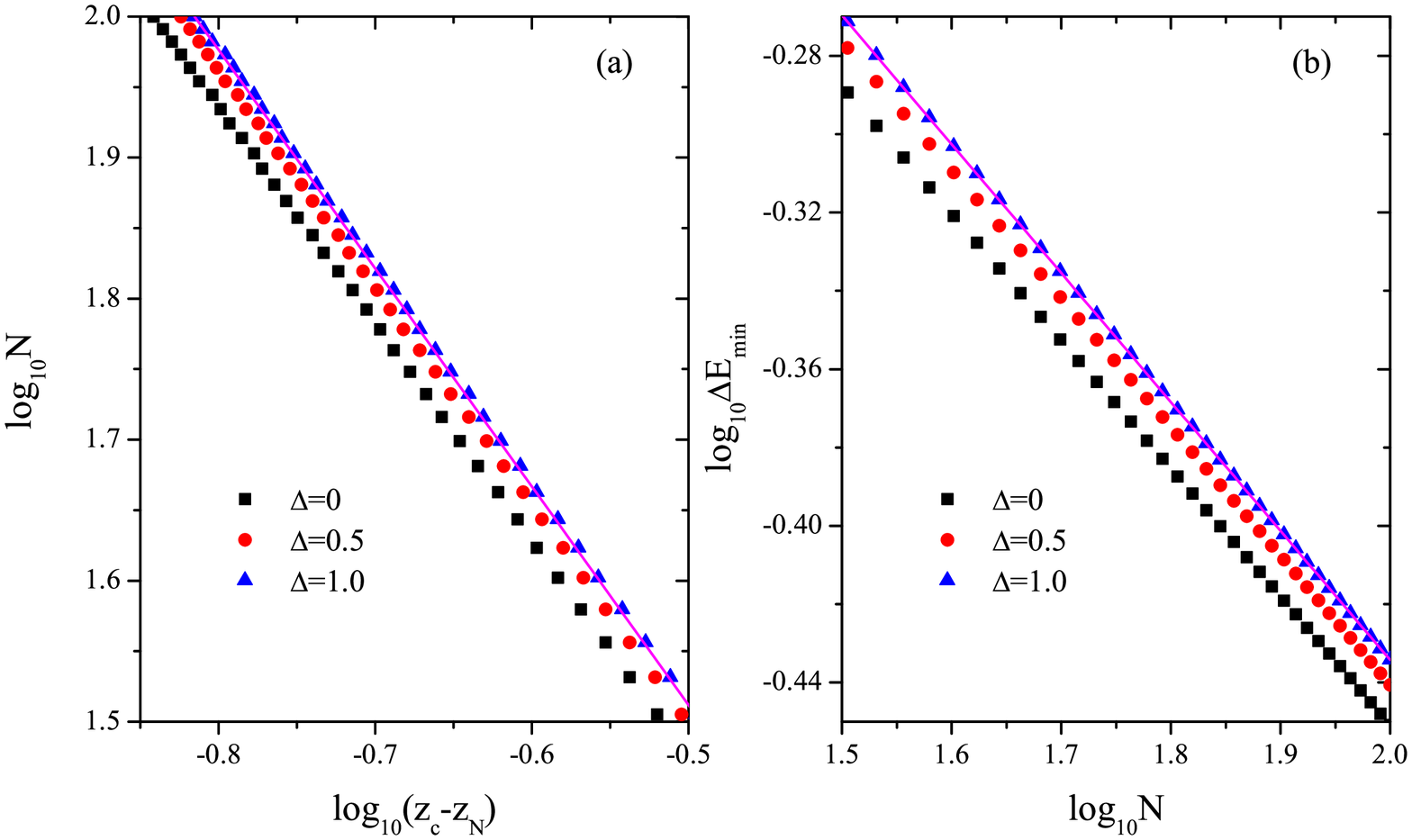}}}
\end{center}
\caption{(Color online) (a) The total atomic number $N$ versus the
offset between $z_N$ and $z_c$. (b) The minimum value of the energy
gap (i.e., $\Delta E_\mathrm{min}$) at the pseudo-critical point
$z_N$ versus $N$. Solid lines are plotted for guiding eyes,
respectively, (from left to right) with the slopes being $-1.54764$
and $-0.32912$.}\label{fig4}
\end{figure}
In order to understand the QPT in the system, we begin our
discussion by analyzing the dimensionless energy gap between the
first excited state and the ground state, namely, $\Delta
E=(E_1-E_0)/\rho$. With the help of diagonalizing the Hamiltonian
(\ref{H2}) numerically, we have calculated the energy levels with
different particle numbers. For a fixed $N$, the energy gap takes a
minimum value at a point $z_N$ (it can be viewed as a
pseudo-critical point of the $N$-particle system) and this point
just corresponds to the position of avoided level crossing (see Fig.
\ref{fig2}). Generally, the QPTs often occur at the positions of
level crossings or avoided level crossings. For our avoided level
crossings system, the existence of the minimum of energy gap
indicates a basic signature of the phase transition. Similar to the
phenomena studied in a two-level atom-molecule system
\cite{santosg10} and in other systems, we find that, the gap $\Delta
E$ in our system also tends to zero at a single point rather than
over an interval of the dimensionless parameter $z/\rho$ with the
particle number $N\rightarrow \infty$. This specific phenomenon
implies that, when $N\rightarrow \infty$, the ground state is
degenerate at the point $z_c$ where is no phase, which is a
requirement for the occurrence of a broken-symmetry phase
\cite{santosg10}.

To capture more features of the QPT in the system, we study the
scaling behavior of the energy gap near the critical point. To this
end, we first calculate the energy gap for different parameter
$\Delta$ and the results have been plotted in Fig. \ref{fig4}.
Either for the variation of the total atomic number $N$ versus
$|z_N-z_c|$ [see Fig. \ref{fig4}(a)] or for the change of the
minimum of the energy gap (i.e., $\Delta E_{\mathrm{min}}$) with
respect to $N$ [see Fig. \ref{fig4}(b)], the same characteristic
scaling laws have been observed for different $\Delta$, and
different lines in each figures with a same slope gives the
evidence.

In the quantum model (\ref{H2}), the total atom number $N$ can be
regard as a correlation length scale of the system, and then one can
connects this length scale to the offset between the pseudo-critical
point and the critical point. Quantificationally, we have
\begin{align}
\kappa|z_c-z_N|^{\nu}\simeq N^{-1},
\end{align}
where $\nu\simeq1.54764$ is a critical exponent and $\kappa\simeq
0.18273$ is a inessential constant. This scaling law shows that the
pseudo-critical point changes and tends as $N^{-1/\nu}$ toward the
critical point and clearly approaches $z_c$ as $N\rightarrow\infty$
[see Fig. \ref{fig4}(a)]. From Fig. \ref{fig4}(b), we find another
scaling law, that is
\begin{align}
\Delta E_\mathrm{min}/N\simeq\Gamma N^{-\zeta},
\end{align}
where $\Gamma\simeq 1.67506$ is a constant. $\zeta\simeq 1.32912$
gives another important exponent, namely, the dynamic critical
exponent. It must be mentioned that all constants and exponents
given in the above two formulas are obtained in the case of
$\Delta=0$, for other cases their values may have a slightly change.
Comparing the product of two exponents in our system with that in a
two-level atom-molecule system \cite{lisc11a} we find that the
values are obviously different. This difference indicates that the
above two models belong to different universality classes.

\subsection{Fidelity}

Similar to other concepts, the behavior of the fidelity can also be
employed to identify the phase transition
\cite{zanardip2006,zhouhq2008}. The fidelity is a measure of the
distance between two states and this concept has been widely used in
the field of quantum information \cite{nielsenma200}.
\begin{figure}[t]
\begin{center}
\rotatebox{0}{\resizebox *{8.5cm}{6.5cm} {\includegraphics
{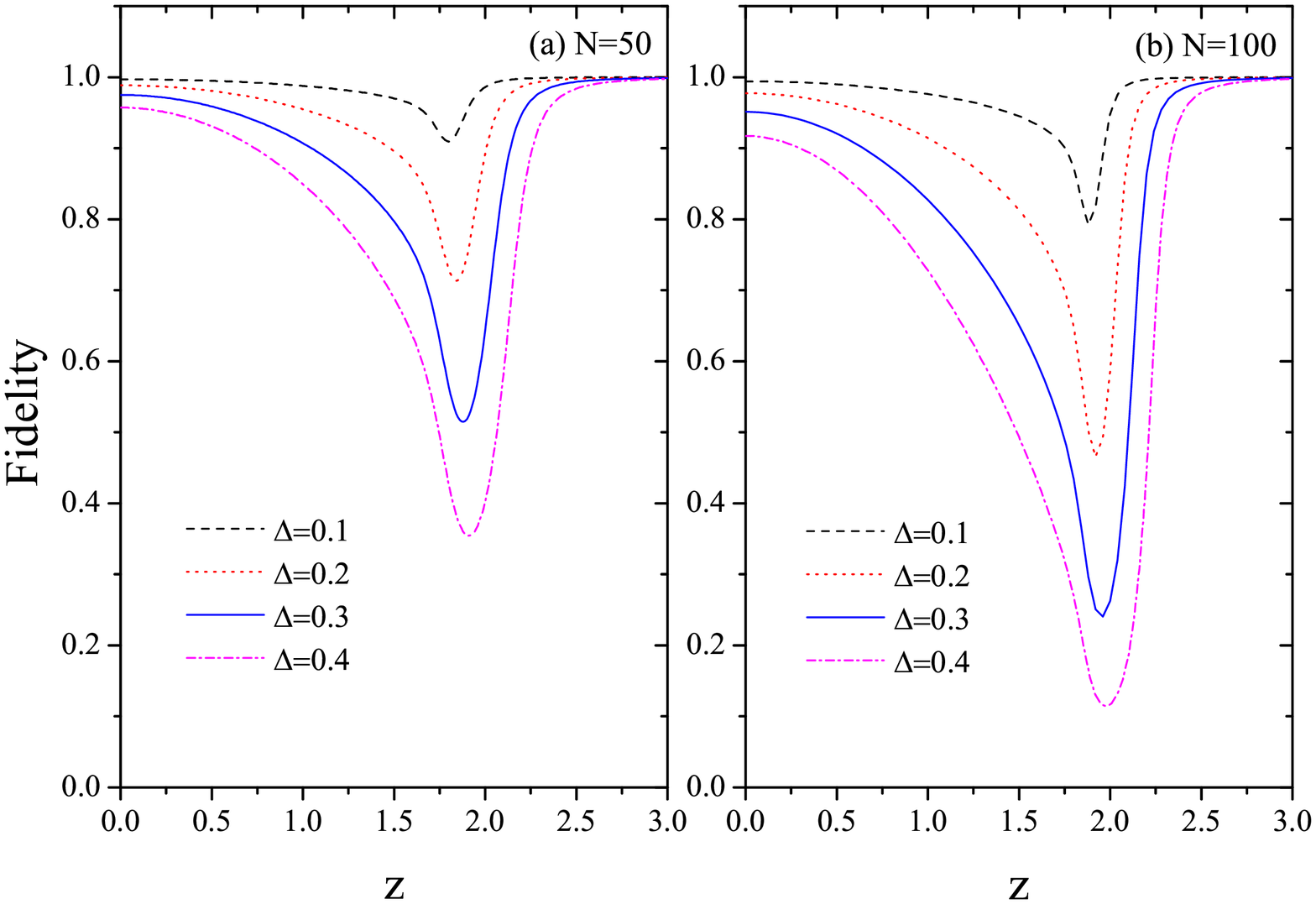}}}
\end{center}
\caption{(Color online) Ground-state fidelity versus the parameter
$z$ with $N=50$ (a) and with $N=100$ (b) for different $\Delta$. The
parameters $\Delta$ and $\rho$ are rescaled by $\rho$.}\label{fig8}
\end{figure}
One can define the fidelity through the modulus of the wave
functions overlap between two states, i.e.,
\begin{align}
F(\psi_1,\psi_2)=|\langle\psi_1|\psi_2\rangle|.
\end{align}
Here we only focus on the behaviors of the fidelity between two
ground states. One ground state is obtained when $\Delta=0$, namely,
$|\Delta=0\rangle$, the other ground state is calculated by treating
$\Delta$ as a perturbation parameter, denoted as $|\Delta\neq
0\rangle$. we have estimated the wave-function overlap between two
ground states with different particle number $N$ and varying
parameter $\Delta$. Figure \ref{fig8} shows the fidelities between
the ground state $|\Delta=0\rangle$ and the ground state
$|\Delta=\alpha\rangle$ with $\alpha=0.1,0.2,0.3$, and $0.4$. Both
for $N=50$ and for $N=100$ we see that the fidelity
$|\langle\Delta=0|\Delta\neq 0\rangle|$ shows dip behavior at the
point corresponding to the pseudo-critical point. The results imply
that the two ground states are distinguishable and there is an
obvious signal for the QPT as long as $\alpha\neq 0$. Moreover, we
observe that the dip of the ground-state fidelity becomes deeper and
the point where the fidelity has a minimum varies with increasing
the value of $\alpha$, this phenomenon is very different from that
in a two-level atom-molecule model where the fidelity has a minimum
at the same point \cite{santosg10}. The reason is that, in our model
the phase transition point $z_c$ is a function of the parameters
$\rho$ and $\Delta$. For a larger value of $\alpha$, we find that
the distinguishability of the two states increases and the minimum
of the fidelity changes evidently.

Now we compare the results obtained in the case of $N=50$ with the
results for $N=100$. For a same $\Delta$, it is seen that the
position of the minimum fidelity for $N=100$ is closer to the
critical point $z_c$ than that for $N=50$. In fact we have
calculated the ground-state fidelity with varying the particle
number $N$ and find that, for a fixed $\alpha$, with increasing $N$,
the wave functions overlap between the two states become smaller
(i.e., the two states are more distinguishable) and the position
where the minimum fidelity occurs moves toward $2\rho+\Delta$. Thus
in the finite particle number case the occurrence of the minima of
the fidelity gives a information about the phase transition in the
system.

\subsection{Geometric phase}

In this subsection we convert to investigate the behavior of the
ground-state geometric phase starting from the mean-field model
(\ref{Hmf}). In the following study we only focus our attention on
the situation that the detuning is absent (i.e., $\Delta=0$).
Because the analytical results can be obtained in this case. To
employ the procedures for calculating geometric phase in nonlinear
systems proposed in Refs. \cite{liuj2010,fulb2010}, we introduce
some new variables through setting
$a=\sqrt{1-2(p_1+p_2)}e^{i\lambda}, b_g=\sqrt{p_1}e^{i
(2\lambda+q_1)}$, and $b_e=\sqrt{p_2}e^{i(2\lambda+q_2)}$, where
$\lambda=\arg(a)$ denotes the total phase, $p_1=|b_g|^2$ and
$p_2=|b_e|^2$ are the population probabilities of the ground state
and excited state molecules, respectively, $q_1=\arg(b_g)-2\arg(a)$
and $q_2=\arg(b_e)-2\arg(a)$ measure the relative phases. With the
help of these new variables, the three-level system can be cast into
a classical Hamiltonian
\begin{align}
\mathcal{H}=&2z\sqrt{p_1p_2}\cos(q_1-q_2)\notag\\&+2\rho\sqrt{p_2}[1-2
(p_1 + p_2)]\cos(q_2+\phi).
\end{align}
The Schr\"{o}dinger equations (\ref{Hmf}) together with the
normalization condition lead to
\begin{align}
\frac{d\lambda}{dt}=-2\rho\sqrt{p_2}\cos(q_2+\phi),\label{lamt}
\end{align}
and
\begin{align}
\frac{dp_i}{dt}=-\frac{\partial\mathcal{H}}{\partial q_i},~~~
\frac{dq_i}{dt}=\frac{\partial\mathcal{H}}{\partial p_i},\label{pqt}
\end{align}
with $i=1,2$. These four equations have established a connection
between the projected Hilbert space spanned by $\mathbf{S}(p_i,q_i)$
and the parameter space spanned by $\mathbf{R}(z,\rho,\phi)$.

Now we calculate the geometric phase for the ground state of the
system. For simplicity, we construct a closed loop $C$ in the
parameter space by treating $z$ and $\rho$ as constants and varying
$\phi$ with time from $0$ to $2\pi$. The system is assumed to evolve
adiabatically along the cyclic path with a rate
$\epsilon\sim|{d\phi\over dt}|\sim{\frac{1}{T}}$, where $T$ is the
period of the cyclic evolution. Initially, we prepare the system in
the ground state of $H_{mf}(\phi=0)$, after a cyclic adiabatic
evolution, the state will acquire a geometric phase besides the
dynamical phase. Here it is noted that the adiabatic parameter
$\epsilon\ll{1}$, and thus it can be regarded as a small parameter
during the process for determining the geometric phase. Following
the method in Ref. \cite{liuj2010}, we expand the total phase
$\lambda$ in a perturbation series in $\epsilon$, i.e.,
\begin{align}\label{lamda}
 {d\lambda\over
 dt}=\lambda_0(\epsilon^0)+\lambda_1(\epsilon^1)+O(\epsilon^2),
\end{align}
to separate the pure geometric part from the total phase. The time
integrals of the zero-order term and the first-order term in Eq.
(\ref{lamda}) will respectively give the dynamic phase and the
geometric phase in the adiabatic limit $\epsilon\rightarrow 0$ or
$T\rightarrow\infty$, and the contribution from the higher-order
terms will vanish.
\begin{figure}[t]
\begin{center}
\rotatebox{0}{\resizebox *{8.5cm}{6.5cm} {\includegraphics
{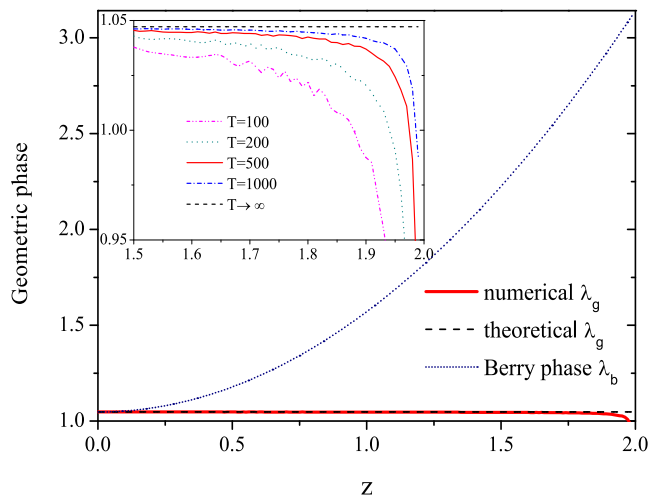}}}
\end{center}
\caption{(Color online) Comparison between the numerical results for
the ground-state geometric phase with $T=500$ and the analytical
results given by Eq. (\ref{berryphase4}). The dotted line denotes
the results obtained from Berry's formula. The inset illustrates the
numerical results around the critical point with different $T$. The
parameter $z$ is rescaled by $\rho$.}\label{fig9}
\end{figure}

During the adiabatic evolution the system will fluctuate around the
ground state due to the small but finite value of $\epsilon$. This
fact allows us to make expansions as
$p_i=\bar{p}_i(\mathbf{R})+\delta{p}_i(\mathbf{R})$ and
$q_i=\bar{q}_i(\mathbf{R})+\delta{q}_i(\mathbf{R})$ with $i=1,2$,
where $(\bar{p}_i,\bar{q}_i)$ stand for the instantaneous ground
state, $\delta p_i$ and $\delta q_i$ denote the fluctuations induced
by the slow change of the system. Substituting these expressions
back into Eq. (\ref{lamt}), when $z<2\rho$, we have
\begin{align}
\lambda_0=&-\mu_0(\mathbf{R}),\label{dynamicphase1}\\
\lambda_1=&{-2
\bar{p}_2\sqrt{\bar{p}_1}z+\rho[\bar{p}_2(6\bar{p}_2+6\bar{p}_1-1)+\delta{p}_2]
\over \sqrt{\bar{p}_2}},
 \label{berryphase1}
\end{align}
where the chemical potential of the ground state can be expressed as
$\mu_0=-2z\sqrt{\bar{p}_1\bar{p}_2}-3\rho\sqrt{\bar{p}_2}[1-2(\bar{p}_1+\bar{p}_2)]$.
Moreover, from Eqs. (\ref{pqt}), we have
\begin{align}\label{berryphase2}
\delta{p}_2=\frac{2\sqrt{\bar{p}_2}(z\bar{p}_2+z\bar{p}_1-4\rho\bar{p}_1^{3/2}
)}{\rho[z(1+6\bar{p}_2+6\bar{p}_1)-16\rho\bar{p}_1^{3/2}]}\frac{d\phi}{dt}.
\end{align}
To deduce Eqs. (\ref{berryphase1}) and (\ref{berryphase2}), we have
used the fixed-point equations
${\partial\mathcal{H}\over\partial{p}_i}|_{(\bar{p}_i,\bar{q}_i)}=0$
and the condition ${d\over dt}\delta{q}_i\sim O(\epsilon^2)$.
Combining Eq. (\ref{berryphase2}) with Eq. (\ref{berryphase1}), and
then using the fixed-point values corresponding to the ground state,
i.e., $(\bar{p}_1,\bar{q}_1;\bar{p}_2,\bar{q}_2)=
(\tfrac{z^2}{16\rho^2},\bar{q}_2\pm\pi;\tfrac{z^2+8\rho^2}{48\rho^2},\pm\pi-\phi)$,
we obtain
\begin{figure}[t]
\begin{center}
\rotatebox{0}{\resizebox *{8.5cm}{6.5cm} {\includegraphics
{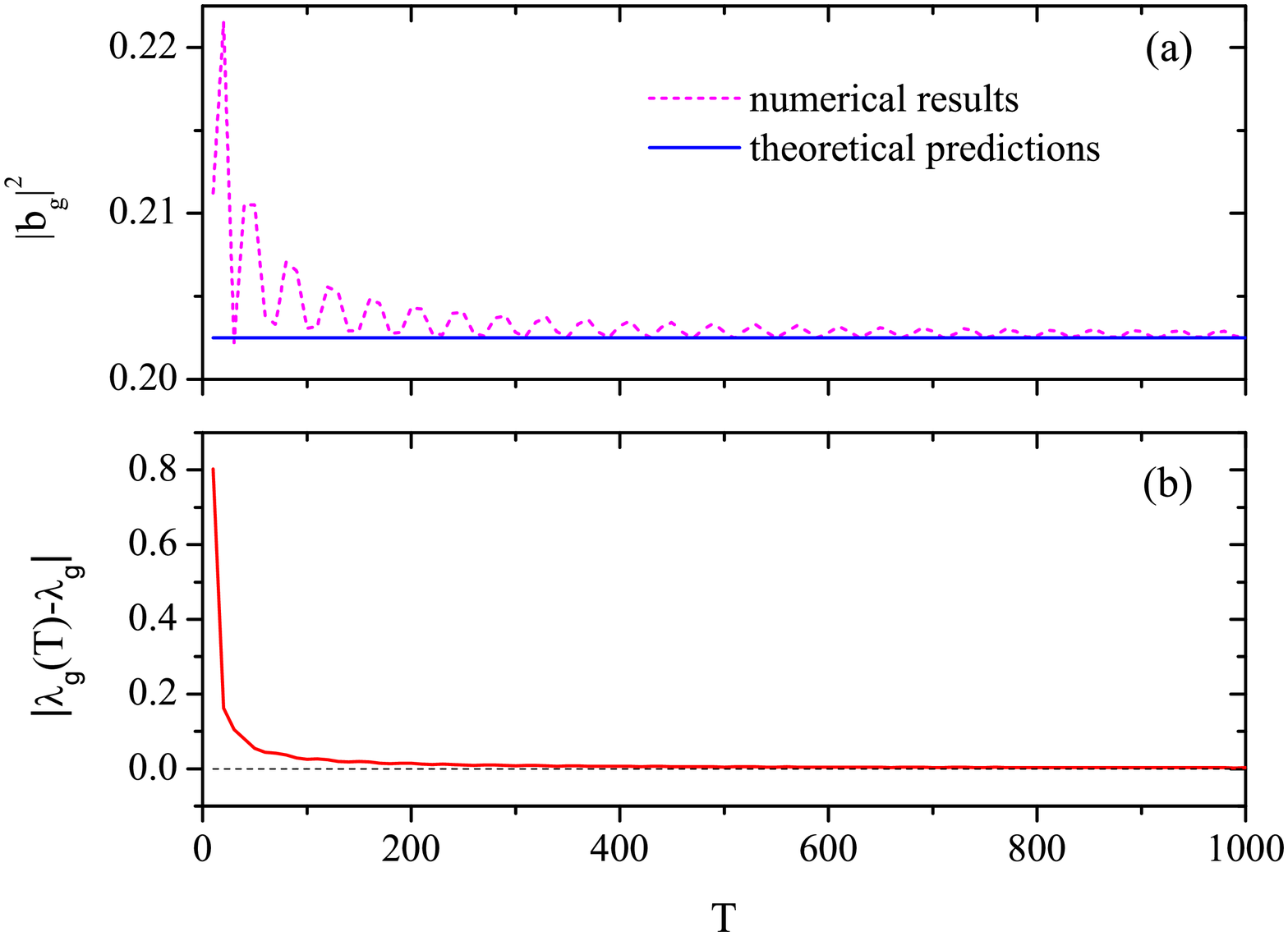}}}
\end{center}
\caption{(Color online) The convergence of population in the
ground-state molecular mode (a) and the geometric phase (b) with
respect to the evolution period $T$ for $z/\rho=1$.}\label{fig10}
\end{figure}
\begin{align}
\lambda_1={1\over 6}{d\phi\over dt}.\label{berryphase3}
\end{align}
By integrating Eq. (\ref{berryphase3}) over $T$ with respect to
time, we have
\begin{align}
\lambda_g=\int_{0}^T\lambda_1dt={1\over
6}\int_0^{2\pi}{d\phi}=\frac{\pi}{3}.\label{berryphase4}
\end{align}
As a comparison we give the result that is directly calculated from
the Berry's formula \cite{berrymv1984}, i.e.,
\begin{align}
\lambda_b=i\int_0^{2\pi}\langle\bar{\psi_0}|\nabla_\phi|\bar{\psi_0}\rangle{d\phi}=\frac{\pi}{6}\left(
2+\frac{z^2}{\rho^2}\right).
\end{align}

Now we consider the situation when $z>2\rho$. In this case, the
ground state, i.e., $\bar{p}_1=\bar{p}_2=1/4$, is independent of the
parameter $\mathbf{R}$. Simple calculations from Eqs. (\ref{lamt})
and (\ref{pqt}) lead to
\begin{align}\label{normalphase}
{d\lambda\over{dt}}={z\over 2}=-\mu_0,
\end{align}
and then
\begin{align}\label{normalphase1}
\lambda=\int_0^T\lambda_0dt=-\int_0^T\mu_0{dt}=\lambda_d.
\end{align}
This result implies that the geometric phase $\lambda_g=0$ in the
case $z>2\rho$. To sum up, our above theoretical calculations
demonstrate that the mean-field geometric phase of the ground state
jumps from zero to $\pi/3$ when the system undergoes the phase
transition from a mixture phase to a pure molecular phase. These
analytical predictions have been checked by numerically solving Eq.
(\ref{Hmf}) or Eqs. (\ref{lamt}) and (\ref{pqt}) as illustrated in
Fig. \ref{fig9}. We see that, if the evolution period $T$ is large
enough, the simulated results show a good agreement with the
analytical predictions. We use Fig. \ref{fig10} to exhibit the
convergence behaviors of the ground-state and the geometric phase
with increasing $T$, and a large convergence rate has been observed.
It is worth emphasizing that, the different values of the geometric
phase in different parameter regions can be an evident signature of
the QPT in the system.

\section{Conclusion}

We have investigated the quantum phase transition in a three-level
atom-molecule conversion system. By using different approaches we
show that the system exists a second-order phase transition similar
to a QPT exhibited in a two-level bosonic model \cite{santosg10}.
Firstly, though analyzing the properties of energy gap we derive two
scaling laws and the corresponding critical exponents. It is found
that the two-level model and our model belong to different
universality classes. Secondly, we discuss the ground-state
fidelity. A minimum value of the fidelity near the critical point
has been found. Finally, we have calculated the ground-state
geometric phase and a discontinues behavior at the critical point
has been observed. This phenomenon is similar to that studied in a
system of a Bose-Einstein condensate in an optical cavity
\cite{lisc11b}. It establishes a connection between the ground-state
geometric phase and the QPT in an interacting atom-molecule bosonic
model following the early works in spin-chain systems
\cite{Carolloacm2005,zhusl2006}. In summary, we have demonstrated
novel characteristic scaling laws and abrupt change of the
ground-state geometric phase in the vicinity of the critical point,
which give the pronounced signals toward the existence of a QPT in
the system.

\section*{ACKNOWLEDGMENTS}
This work is supported by the National Fundamental Research Program
of China (Contract No. 2011CB921503) and the National Natural
Science Foundation of China (Contracts No. 10725521, No. 91021021,
No. 11075020, and No. 11078001).


\begin{thebibliography}{99}
\bibitem{sachdevs2003}S. Sachdev, Quantum Phase Transitions (Cambridge University
Press, Cambridge, 1999); M. Vojta, Rep. Prog. Phys. {\bf 66}, 2069
(2003).
\bibitem{sondhisl1997} S. L. Sondhi, S. M. Girvin, J. P. Carini, and D. Shahar, Rev.
Mod. Phys. {\bf 69}, 315 (1997).
\bibitem{greinerm2002}M. Greiner, O. Mandel, T. Esslinger, T. W.
H\"{a}nsch, and I. Bloch, Nature {\bf 415}, 39 (2002).
\bibitem{tilahund2011}D. Tilahun, R. A. Duine, and A. H. MacDonald, Phys. Rev. A
{\bf 84}, 033622 (2011).
\bibitem{pitaevskiil2003}See, e.g., L. Pitaevskii and S. Stringari, {\it Bose-Einstein Condensation}
 (Clarendon Press, Oxford, 2003).
\bibitem{ruseckasj2005}J. Ruseckas, G. Juzeli\={u}nas, P. \"{O}hberg, and M. Fleischhauer,
Phys. Rev. Lett. {\bf 95}, 010404 (2005).
\bibitem{Donleyea2002}E. A. Donley, N. R. Claussen, S. T. Thompson, and C. E. Wieman,
Nature (London) {\bf 417}, 529 (2002).
\bibitem{xuk2003} K. Xu, T.Mukaiyama, J. R. Abo-Shaeer, J. K. Chin, D. E. Miller, and W. Ketterle, Phys. Rev.
Lett. {\bf 91}, 210402 (2003).
\bibitem{Herbigj2003} J. Herbig, T. Kraemer, M. Mark, T. Weber, C. Chin, H.-C. N\"{a}gerl, and R. Grimm, Science
{\bf 301}, 1510 (2003).
\bibitem{Wynarr2000} R. Wynar, R. S. Freeland, D. J. Han, C. Ryu, and D. J. Heinzen,
Science {\bf 287}, 1016 (2000).
\bibitem{Romt2004}T. Rom, T. Best, O. Mandel, A. Widera, M. Greiner, T. W. H\"{a}nsch,
and I. Bloch, Phys. Rev. Lett. {\bf 93}, 073002 (2004).
\bibitem{whinklerk2005}K. Winkler, G. Thalhammer, M. Theis, H. Ritsch, R. Grimm, and J. H.
Denschlag, Phys. Rev. Lett. {\bf 95}, 063202 (2005).

\bibitem{Gaubatzu1988}U. Gaubatz, P. Rudecki, M. Becker, S. Schiemann, M. K\"{u}lz, and K.
Bergmann, Chem. Phys. Lett. {\bf 149}, 463 (1988).
\bibitem{Kuklinskijr1989} J. R. Kuklinski, U. Gaubatz, F. T. Hioe,
and K. Bergmann, Phys. Rev. A {\bf 40}, 6741 (1989).
\bibitem{Pazye2005}E. Pazy, I. Tikhonenkov, Y. B. Band, M. Fleischhauer, and A. Vardi,
Phys. Rev. Lett. {\bf 95}, 170403 (2005).
\bibitem{linghy2007}H. Y. Ling, P. Maenner, W. Zhang, and H.
Pu, Phys. Rev. A {\bf 75}, 033615 (2007).
\bibitem{Shapiroea2007}E. A. Shapiro, M. Shapiro, AviPe'er, and J. Ye,
Phys. Rev. A {\bf 75}, 013405 (2007).
\bibitem{mengsy08}S.-Y. Meng, L.-B. Fu, and J. Liu, Phys. Rev. A {\bf 78}, 053410 (2008).
\bibitem{fulb2010}L. B. Fu and
J. Liu, Ann. Phys. {\bf 325}, 2425 (2010).
\bibitem{wub2011}F. Cui and B. Wu, Phys. Rev. A {\bf 84}, 024101
(2011).
\bibitem{santosg10} G. Santos, A. Foerster, J. Links, E. Mattei, and S. R.
Dahmen, Phys. Rev. A {\bf 81}, 063621 (2010).
\bibitem{lisc11a}S. C. Li and L. B. Fu, Phys. Rev. A {\bf 84}, 023605
(2011).


\bibitem{puh07}H. Pu, P. Maenner, W. Zhang, and H. Y. Ling, Phys.
Rev. Lett. {\bf 98}, 050406 (2007).
\bibitem{linghy04}H. Y. Ling, H. Pu, and
B. Seaman, Phys. Rev. Lett. {\bf 93}, 250403 (2004).
\bibitem{linghy07}H. Y. Ling,
P. Maenner, W. Zhang, and H. Pu, Phys. Rev. A {\bf 75}, 033615
(2007).



\bibitem{regalca2004} C. A. Regal, M. Greiner, and D. S. Jin, Phys. Rev. Lett. {\bf 92},
040403 (2004); G. B. Partridge, K. E. Strecker, R. I. Kamar, M. W.
Jack, and R. G. Hulet, {\it ibid}. {\bf 95}, 020404 (2005).
\bibitem{linksj2003}J. Links, H. Q.
Zhou, R. H. McKenzie, and M. D. Gould, J. Phys. A {\bf 36}, R63
(2003); G. Santos, A. Tonel, A. Foerster, and J. Links, Phys. Rev. A
{\bf 73}, 023609 (2006); J. Li, D. F. Ye, C. Ma, L. B. Fu, and J.
Liu, {\it ibid}. {\bf 79}, 025602 (2009).

\bibitem{zanardip2006}P. Zanardi and N. Paunkovi\'{c}, Phys. Rev.
E {\bf 74}, 031123 (2006).
\bibitem{zhouhq2008}H.-Q. Zhou and J. P. Barjaktarevic,
J. Phys. A {\bf 41}, 412001 (2008).
\bibitem{nielsenma200}M. A. Nielsen and I. L. Chuang,
{\it Quantum Computation and Quantum Information} (Cambridge
University Press, Cambridge, UK, 2000).
\bibitem{liuj2010} J. Liu and L. B. Fu, Phys. Rev. A {\bf 81}, 052112 (2010).

\bibitem{berrymv1984}M. V. Berry, Proc. R. Soc. London, Ser. A {\bf 392}, 45 (1984).

\bibitem{lisc11b}S. C. Li, L. B. Fu, and J. Liu, Phys. Rev. A {\bf 84}, 053610
(2011).
\bibitem{Carolloacm2005}A. C. M. Carollo and J. K. Pachos, Phys. Rev. Lett. {\bf 95}, 157203 (2005).
\bibitem{zhusl2006}S.-L. Zhu, Phys. Rev. Lett. {\bf 96}, 077206 (2006).
\end{thebibliography}
\end{document}